%% file: main.tex
\documentclass[%
 reprint,
 amsmath,amssymb,
 aps,
]{revtex4-2}

\usepackage{graphicx}
\usepackage{dcolumn}
\usepackage{bm}
\usepackage{booktabs}
\usepackage{graphicx}

\usepackage{color}
\usepackage{xcolor}

\usepackage{array}
\usepackage{makecell}

\usepackage[english
]{babel}

\begin{document}

\title{\textbf{Ionization-free femtosecond UV pulse filamentation resulting from non-perturbative Kerr effect saturation due to transient photoexcitation of molecules}}%

\author{
V.~V.~Strelkov$^{\ast}$, A.~V.~Shutov,  V.~D.~Zvorykin
}
\affiliation{
\mbox{P.N. Lebedev Physical Institute of the Russian Academy of Sciences, 53 Leninskiy Prospekt,  Moscow 119991, Russia} \\
$^{\ast}$strelkov.v@gmail.com
}

\date{\today}

\begin{abstract}
Our experimental and numerical study of filamentation of the 248 nm 100 fs pulses in air, nitrogen, and oxygen at atmospheric pressure shows that the densities of photoelectrons in the filaments are not sufficient to limit Kerr self-focusing. The simulations explain this result, showing that the nonlinear refractive index growth with the UV intensity is saturated due to transient photoexcitation of a molecule, not the photoionization. Correct simulation of the observed filament intensity ratio in different gases validates the suggested mechanism of the femtosecond UV pulses filamentation.
\end{abstract}

\maketitle



Laser pulses with femtosecond or picosecond duration and gigawatt power, launched in air or other gases, form self-guided light channels propagating over many Rayleigh lengths. These filaments originate from an interplay of self-focusing due to the optical Kerr effect, and some mechanism switching on at high intensities and preventing further self-focusing. Within the so-called standard model of filamentation~\cite{Brodeur_1997, Mlejnek_1998, Couairon_2000, Berge_2007, Couairon_2007, Kosareva_2011, Chin_2012, Kandidov_2013}, this limiting mechanism is multiphoton photoionization. Ionization occurs mainly in the region of the highest intensities, i.e. near the beam axis, and the negative contribution to the refractive index due to free-electrons \textcolor{purple}{provides defocusing. When this plasma defocusing  compensates the Kerr focusing filamentation occurs}. Other limiting mechanisms were suggested and studied in~\cite{Mechain_2004, Loriot_2009, Bejot_2010, Ettoumi_2010}.  The dominating process that limits self-focusing depends on the pulse wavelength, its duration, and the type of the gas.

Studies based on numerical time-dependent Schr\"odinger equation (TDSE) solution allow investigation both saturation of the Kerr nonlinearity of the bound electrons and the role of the free electrons~\cite{Nurhuda_2008,  Ettoumi_2010, Teleki_2010, Bree_2011, Volkova_2011, Volkova_2012, Skupin_2013, Bejot_2013}. However, for the IR field, even if saturation of the Kerr nonlinearity takes place, it occurs at high field intensities and thus unavoidably accompanied by gas photoionization.  Refs.~\cite{Volkova_2012, Skupin_2013} conclude that free electrons play a dominant role in self-focusing compensation in the IR field. 

The study of filamentation for the UV pulses has started somewhat later than for the IR ones, but up to date UV filaments have been observed and investigated in many experiments~\cite{Schwarz_2000, Tzortzakis_2001, Couairon_2002, Daigle_2010, Smetanin_2016, Shipilo_2017, Shipilo_2017a, Zvorykin_2020, Rastegari_2021}. The intensity in the UV filament and the plasma density are similar~\cite{Couairon_2002, Daigle_2010} or lower~\cite{Tzortzakis_2001, Smetanin_2016, Zvorykin_2020} than in the IR case. These observations are hardly compatible with the standard filamentation model because the free-electrons' contribution to the refractive index decreases with the fundamental frequency $\omega_0$ (as $1/\omega_0^2$ \textcolor{purple}{ in line with the Drude model for low collision frequency}) and the Kerr nonlinearity increases {\color{purple} (at least for air, nitrogen, and oxygen) in the considered frequency range}. Using published data~\cite{Shaw1993, Nibbering_1997, Loriot_2009} for the latter nonlinearity, one finds that the balance between the lowest-order Kerr (in air) and plasma contributions to the refractive index requires about 43 times higher electron concentration for the 248~nm pulse than for 800~nm pulse of the same intensity. Such high electron concentrations not only contradict the reported ones, but would also lead to pronounced UV absorption which is not compatible with the observed UV filaments with length of several meters (or even longer~\cite{Shipilo_2017, Shipilo_2017a}). Moreover, when the fundamental frequency changes from IR to near-UV, the percentage of molecular excitation increases~\cite{Daigle_2010, Volkova_2011, Vrublevskaya_2023}, potentially leading to modification of the optical properties of the gas. In general, one can conclude that for UV filaments, the dominating mechanism limiting self-focusing is not yet clear.   

 In this paper, we experimentally study the filamentation of 248~nm femtosecond pulses in air, nitrogen, and oxygen. Moreover, we simulate the nonlinear refractive index in the intense UV field using the numerical solution of the 3D time-dependent Schr\"odinger equation (TDSE). The observed ratio of the filament intensities in different gases is well reproduced in our calculations. Thus validated, the simulation allows us to clearly identify \textcolor{purple}{that the transient molecular excitation is} the mechanism that limits self-focusing in the UV filament for pulse durations of about 100 fs. 

Collimated ultraviolet (UV) {\color{purple} filaments in nitrogen, oxygen, and dry air} of atmospheric pressure are studied experimentally using propagation in a gas pumped through a tube with one entrance open for the incoming light beam to eliminate artifacts which would have been caused by a window in case of using a sealed gas cell. 
The experimental schematic is shown in Fig. 1a. Frequency tripled Ti:Sapphire laser~\cite{Zvorykin_2017} delivered 248 nm 100 fs long pulses with an energy of approximately E=200~\textemdash~300 $\mu$J. The collimated laser beam {\color{purple} passed several meters in air (this length is controlled by the movable mirror position) and then entered the open end of a 4-meter long 25-mm inner diameter tube. The length of the tube was limited by the available space in our laboratory, and the initial propagation in the air provided the starting stage of filament formation.} A K8 glass window, opaque to 248-nm UV radiation, was mounted on the tube's exit. This output window served as a luminescent screen that converted UV radiation into visible. The fluorescence pattern from the window was imaged onto an Ophir SP-620U CCD camera using a macro objective lens. The luminescent response of the K8 glass plate was calibrated against the pulse energy (measured by the calorimeter) using the same laser establishing a direct relationship between the luminescence signal recorded by the CCD and the incident laser intensity profile (see~\cite{Zvorykin_2020} for more details). The pulse energy was controlled by Ophir Nova II calorimeter for each position of the movable mirror. The Reynolds number for the flow was estimated to be between 500 and 1000. As this is well below the critical value of $\sim 2500$ for air, O$_2$, or N$_2$ in a smooth pipe, the gas flow was considered laminar. Filamentation was typically observed at the total distance of 6~\textemdash~9 meters from the laser.

The normalized transverse profiles of the UV beam after propagation are presented in Fig. 1b for different gases and propagation lengths. {\color{purple} We see that the intensity profiles are similar for different propagation lengths (except the shortest one for oxygen) confirming the filament formation.} The pulse energy was measured by the calorimeter for each propagation length shown in Fig 1b. The measured average {\color{purple}(excluding the results for the shortest propagation length)} filament diameters and intensities $I_{exp}$ calculated using these diameters, the pulse energy and duration are summarized in Table 1. The intensity measured for oxygen is about 3 times {\it higher} than for air and nitrogen. Within the filamentation picture where the plasma defocusing limits the intensity, this is a counterintuitive result because the oxygen's ionization energy is lower (in particular, three and four photons are necessary to ionize the oxygen and nitrogen molecule, respectively) so one should expect {\it lower} intensity to compensate for self-focusing. The quantitative estimation of the free-electron densities in filaments shows further disagreement with this picture. Using the filament intensities and the ionization cross-sections for nitrogen and oxygen measured in~\cite{Shutov_2017, Shutov_erratum}  we calculate the peak electron densities $N_e$. Finally, using nonlinear refractive indexes~\cite{Shaw1993, Nibbering_1997} $n_2$ we calculate the parameters $n_2 I_{exp}$ and $N_e / 2N_{cr} $, where $N_{cr}$ is the critical plasma density for the 248 nm radiation. The latter two parameters define the role of Kerr self-focusing and plasma defocusing, respectively~\cite{Couairon_2007}. We see that the free-electron density, although very different in different gases, is orders of magnitude lower than what is required for the self-focusing compensation. Despite some uncertainty of the intensity measurement, the huge ratio of the parameters $n_2 I_{exp}$ and $N_e / 2 N_{cr}$ allows us to conclude that plasma defocusing is not the mechanism that limits Kerr self-focusing. This encourages us to investigate other mechanisms. 

\begin{figure}
a)

\includegraphics[width=0.5\textwidth]{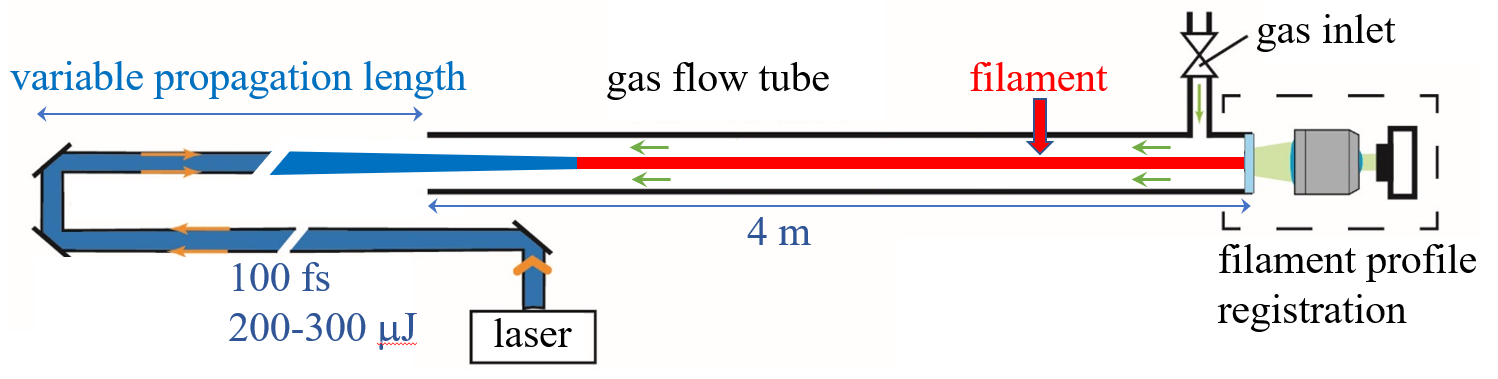}
b)

\includegraphics[width=0.4\textwidth]{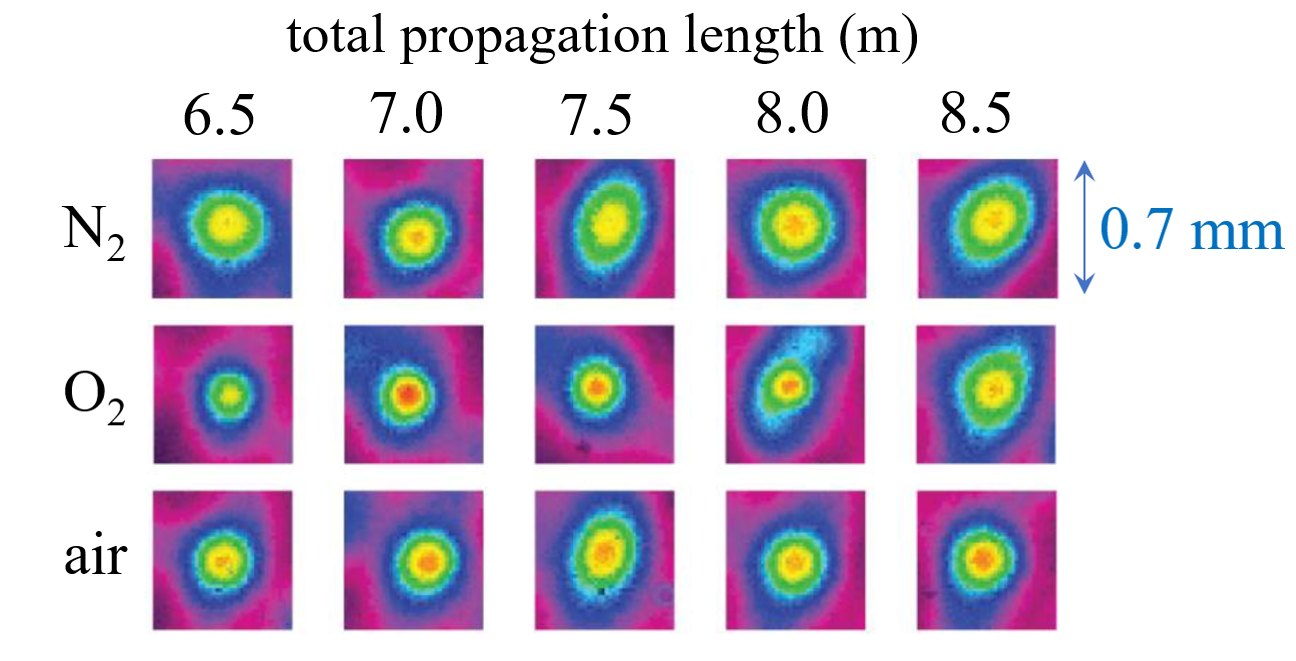}
\caption{(a) Experimental set-up. (b) Measured transverse UV filament profiles in N$_2$, O$_2${\color{purple}, and air} for different propagation distances.}
\label{Fig_exp}
\end{figure}

\begin{table}
\caption{Filament parameters for different gases: experimental filament diameter $d$, experimental intensity $I_{exp}$ in the center of the filament, estimated electron density $N_e$, published nonlinear refractive index $n_2$, , as well as products of the parameters, defining {\color{purple} the contributions to the refractive index due to the Kerr effect $n_2 I_{exp}$ and free-electrons $\frac{N_e}{2 N_{cr}}$. Theoretical filament intensity $I_{th}$ is calculated via Eq.~(\ref{I_fil}), see text for more details.}}
{\color{purple}
  \begin{center}
      \resizebox{0.53\textwidth}{!}{%
    \begin{tabular}{ | c | c | c| c| c| c| c | c|}
    \hline
    \thead{gas} &  \thead{d, \\ $\mu$m} & \thead{$I_{exp}$, \\ TW/cm$^2$ } & \thead{$N_e$,  \\ $10^{12}$~cm$^{-3}$}  & \thead{$n_2$, \\cm$^2 /$W} & \thead{$n_2 I_{exp}$ \\ \quad}   & \thead{$\frac{N_e}{2 N_{cr}}$ \\ \quad }  & \thead{$I_{th}$, \\ TW/cm$^2$ } \\
    \hline
    \thead{N$_2$} & 300$\pm$30 & 0.23$\pm$0.06 & 0.05  & 7.5 10$^{-19}$ & 1.7 10$^{-7}$ &  1.4 10$^{-12}$ & \shortstack{TDSE: 0.34 \\ analytic: 0.17}  \\
    \thead{O$_2$} & 220$\pm$30 & 0.7$\pm$0.2 & 10.3  & 3.0 10$^{-18}$ & 2.1 10$^{-6}$ &  2.9 10$^{-10}$ & TDSE: 1.0 \\
   \thead{air}   & 260$\pm$30 & 0.33$\pm$0.09 & 0.35  &   1.2 10$^{-18}$ & 4.0 10$^{-7}$ &   1.0 10$^{-11}$ & TDSE: 0.70 \\
    \hline
    \end{tabular}}
  \end{center}
  \label{Table_exp}
  }
  \end{table}

In our simulations the nonlinear refractive index is found by numerically solving the 3D time-dependent Schr\"odinger equation (TDSE) for a model system in a laser field. \textcolor{purple}{Under the specific sets of the model potential's parameters it partly reproduces the structure of the nitrogen or the oxygen molecule's energy levels, see more details in Supplement A.} Solving TDSE for the laser pulse with duration $\tau$ and peak intensity $I$ we find the wave function $\Psi(I, \tau, t)$, $t$ is the time instant.  Then we calculate the microscopic polarization response and find its spectrum $d(I, \tau, \omega)$. The polarizability is calculated as
\begin{equation}
\alpha (I, \tau)=d(I, \tau, \omega_0)/ E(I, \tau,  \omega_0),
\label{polarizability}    
\end{equation}
where $E(I, \tau, \omega)$ is the spectrum of the laser field, $\omega_0$ is the fundamental frequency. Moreover, we calculate the excitation probability at time instant $t$ as $p_{ex}(I, \tau,t)=1-|<\psi_0|\Psi(I, \tau, t)>|^2$, $\psi_0$ is the ground state wave-function. Then the {\it nonlinear} excitation probability  is calculated as:
\begin{equation}
p_{ex}^{(nl)} (I,\tau,t)=p_{ex}(I, \tau,t)-p_{ex}(I^{(weak)}, t) I/I^{(weak)},
\label{ex_nl}    
\end{equation}
where $I^{(weak)}$ is a certain value of low intensity, for which the linear excitation firmly dominates. In the calculations, we use $I^{(weak)}=0.1$~TW/cm$^2$. {\color{purple} Note that the nonlinear excitation of a molecule by UV fields plays a key role in resonance-enhanced multiphoton ionization (REMPI)~\cite{Ogorzalek_1989, Guthrie_1990, Trickl_1993, Mori_1999, McGuire_2014, Gardner_2019} and air lasing~\cite{Nie_2024}. 
}

\begin{figure}
\includegraphics[width=0.5\textwidth]{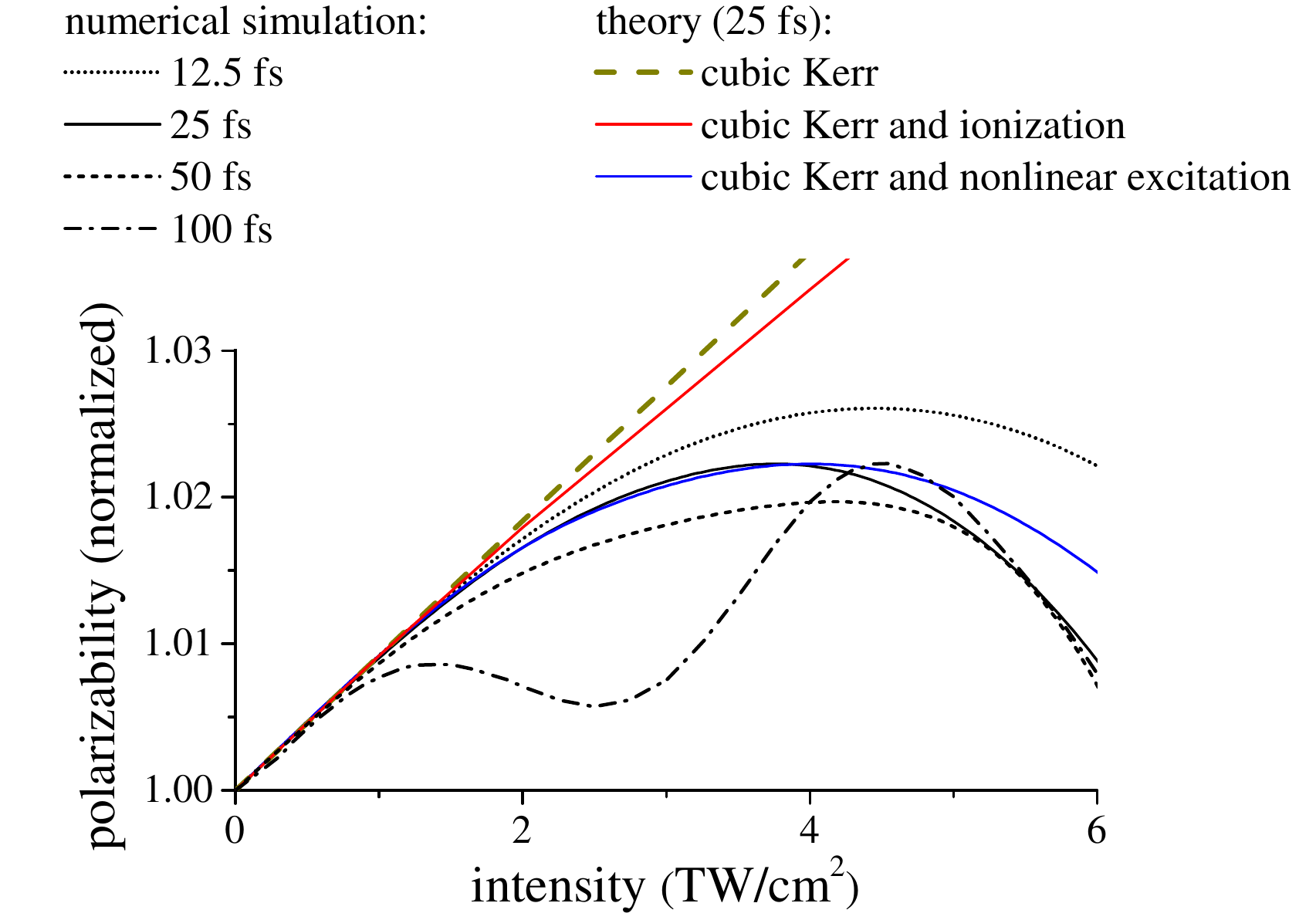}
\caption{The normalized polarizability $\alpha/\alpha_0$ calculated numerically via Eq.~(\ref{polarizability}) \textcolor{purple}{for nitrogen} for different pulse durations (black curves), and analytically for a 25 fs long pulse taking into account only the lowest-order (i. e. cubic) Kerr effect (dashed dark yellow line), the cubic Kerr effect and ionization (solid red curve), the cubic Kerr effect and nonlinear nonperturbative photo-excitation via Eq.~(\ref{polarizability_app}) (solid blue curve).}
\label{al_vs_int}
\end{figure}

The polarizability \textcolor{purple}{calculated for nitrogen} as a function of the peak laser intensity is shown in Fig.~\ref{al_vs_int}. For low intensities it grows linearly with the laser intensity due to the cubic Kerr effect. However, at a certain intensity, it saturates and then rapidly decreases. The saturation intensity decreases with pulse duration varying from approximately 4 TW/cm$^2$ for 50 fs and shorter pulses down to approximately 1.5 TW/cm$^2$ for a 100 fs pulse. {\color{purple}Similar values of the saturation intensity are found for certain range of UV frequencies, see Appendix F.}  These values are close to those found in the TDSE simulations for UV pulses~\cite{Vrublevskaya_2023} (although without separating the contributions of atoms and free electrons) whereas for 800~nm radiation the simulations demonstrate the saturation of the Kerr effect  for much higher intensities (order of magnitude higher~\cite{Ettoumi_2010, Bree_2011, Teleki_2010, Vrublevskaya_2023} or even more~\cite{Nurhuda_2008}).

Note that the fact that for high intensities the nonlinear polarizability depends not only on the intensity but also on the pulse duration is a clear evidence of the nonperturbative character of this nonlinearity~\cite{ABecker2017}.

The figure also presents the linear approximation based on the polarization behavior for low intensities $\alpha_0+\alpha_2 \ I$ (denoted as "cubic Kerr" in the graph), and the sum of this term and the contribution of free electrons to the polarizability, see Supplement A for the calculation details. One can see that the latter contribution is negligible (in agreement with our experiments) and cannot explain the saturation and the decrease of the polarizability with the laser intensity. In contrast, the following equation describes this behavior very well:
\begin{equation}
\alpha_{app}(I, \tau)=(\alpha_0+\alpha_2 \ I) \left(1-\bar{p}_{ex}^{(nl)} (I, \tau)\right)+\alpha_{el}  \ \bar{p}_{ex}^{(nl)} (I, \tau),
\label{polarizability_app}    
\end{equation}
here $\bar {p}_{ex}^{(nl)} (I, \tau)=p_{ex}^{(nl)} (I, \tau,t=\tau/2)$ is the nonlinear excitation probability (given by Eq.~(\ref{ex_nl})) in the center of the pulse, the polarizability of the free electron is (in atomic units) $\alpha_{el}=- 1/ \omega_0^2$. In Eq.~(\ref{polarizability_app}) the first term describes the decrease of the refractive index due to (the nonlinear) depopulation of the ground state. The second term describes the polarizability of the (nonlinearly) populated excited states: we roughly assume that the polarizability of the excited molecule is equal to that of the free electron. \textcolor{purple}{The numerical TDSE calculations in Appendix E show reasonable applicability of this approximation for the UV field in our conditions. In general, the origin of this applicability is the fact that the excited molecule} finds itself in a state for which the lowest transition frequency for further excitation is {\it below} the UV field's frequency, see Fig.~S1 in Supplement A. In our conditions, the main contribution to the deviation from the cubic Kerr effect is provided by the first term in Eq.~(\ref{polarizability_app}). Note that a similar equation was suggested to describe the nonlinear polarization response in the atomic stabilization regime in intense laser field~\cite{volkova_nonlinear_2013}.  

\begin{figure}
\includegraphics[width=0.4\textwidth]{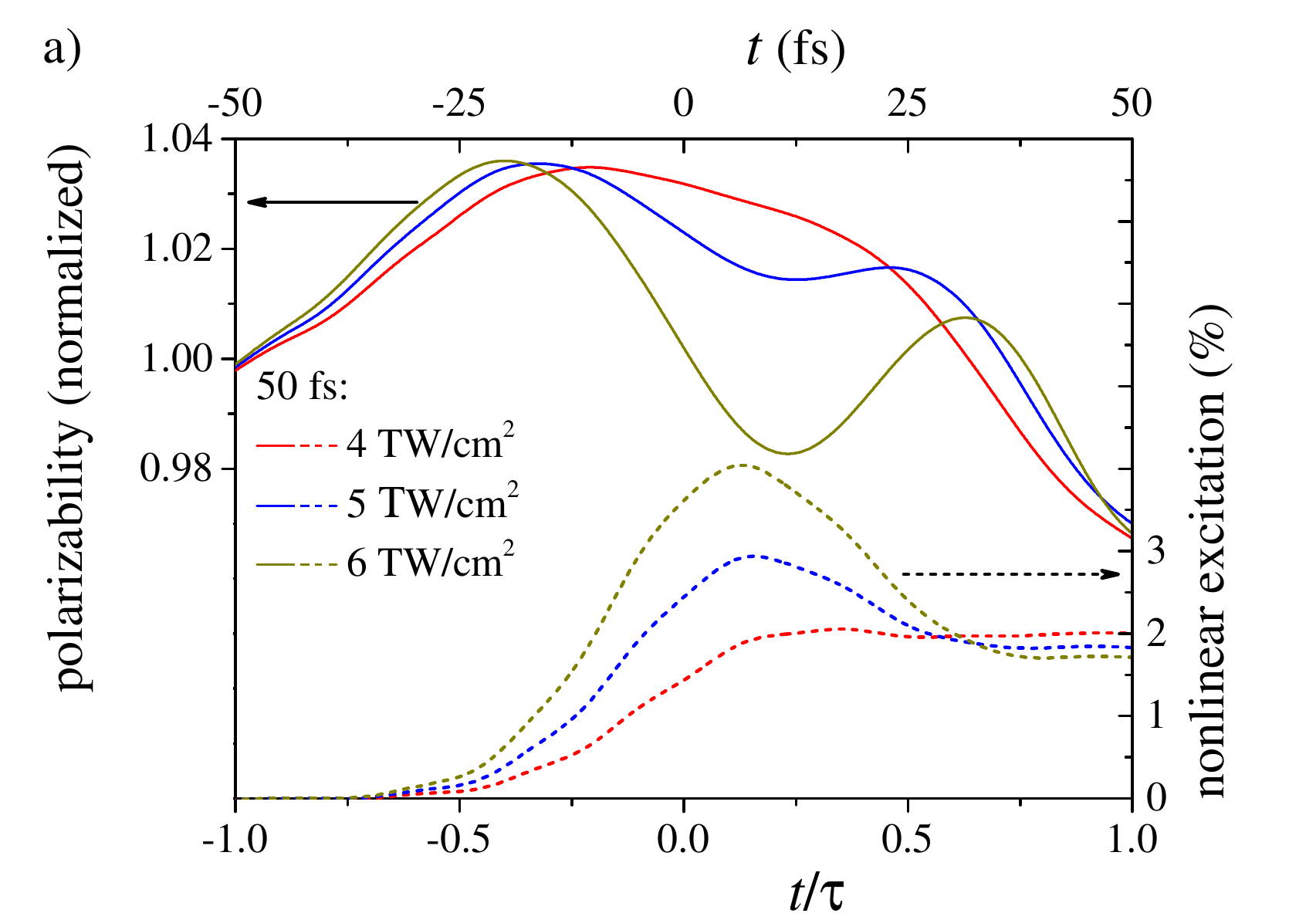}
\includegraphics[width=0.4\textwidth]{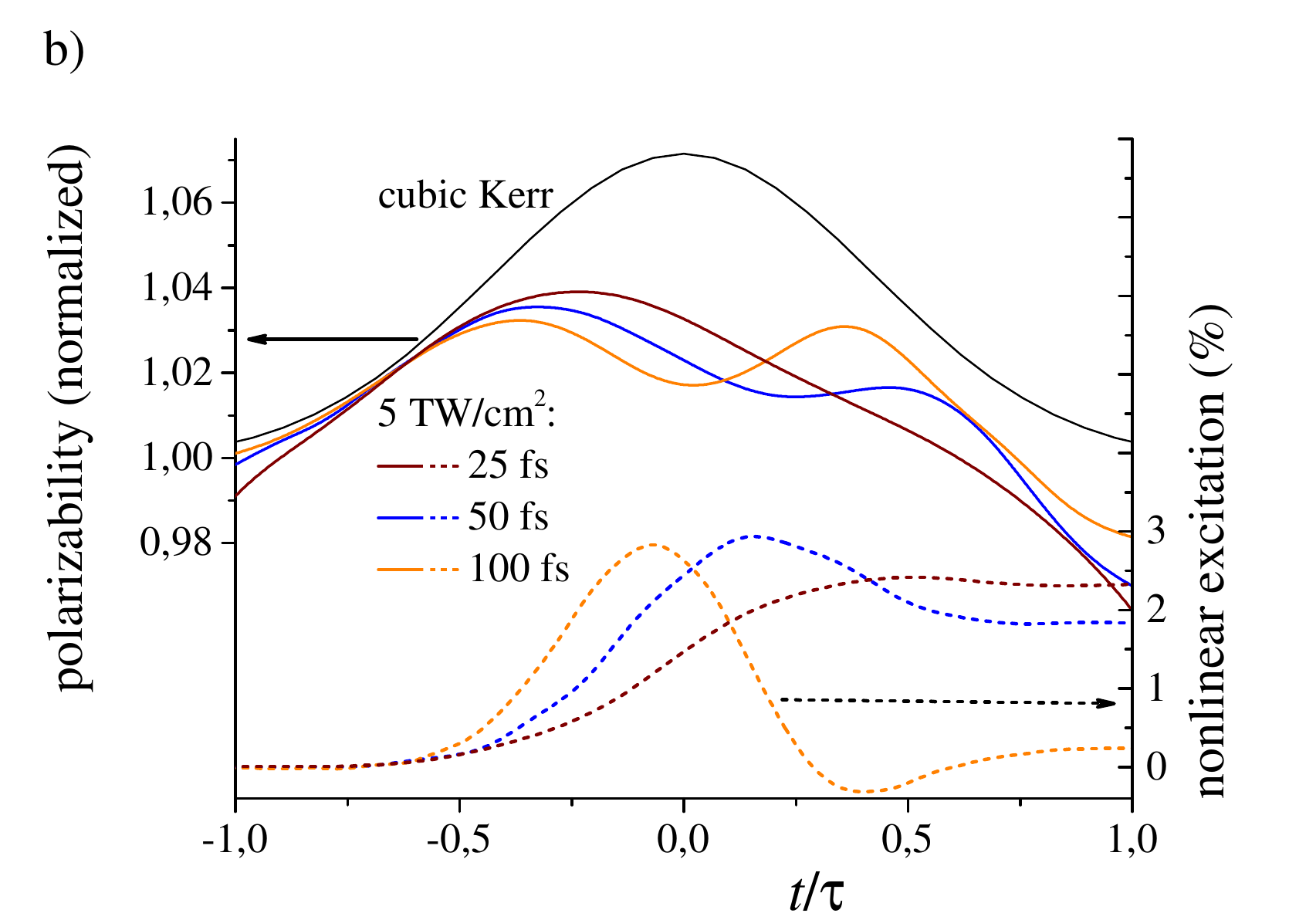}
\caption{The normalized transient polarizability (solid curves) and nonlinear excitation (dashed curves) as functions of time calculated for different peak pulse intensities (a) and different pulse durations (b) \textcolor{purple}{for nitrogen}.}
\label{Gabor}
\end{figure}

In Fig.~\ref{Gabor} we present the transient nonlinear excitation calculated via Eq.~(\ref{ex_nl}) together with the time-dependent polarizability. The latter is calculated using wavelet transforms of the polarization and field, see Supplement B for more details. One can clearly see that the difference between the actual polarizability and the Gaussian dependence expected from the cubic Kerr-effect follows the transient nonlinear excitation. Moreover, for higher intensities and durations, the transient excitation within the pulse can be higher than the residual excitation after the pulse, see Supplement C for more details. So, the laser field absorption, defined by the {\it residual} excitation, is not very high.

Note, that for the high intensities and pulse durations in Fig.~~\ref{Gabor} we see the oscillations of the polarizability as a function of time. Similar oscillations were found in non-perturbative analytical studies of the two-photon resonance~\cite{Belenov_1992, Isakov_1996, Zvorykin_2014}. 

The saturation of the nonlinear refractive index limits the intensity in the filament and our numerical results allow estimating the intensity in the filament for a given beam power as described below. For relatively low intensities when the deviation from the linear dependency is not very high, one can approximate the refractive index as:
\begin{equation}
 n(I)=n_0+n_2I-n_4 I^2.   
 \label{n_full}
\end{equation}
\textcolor{purple}{Supplement D describes the applicability range of this approximation. The refractive index given by Eq.~(\ref{n_full}) stops growing under $I_{sat}=n_2/(2 n_4)$. Below we denote this intensity as the saturation intensity.}

Let us introduce an effective intensity-dependent nonlinear refractive index 
\begin{equation}
n_2^{(eff)} (I) \equiv \partial n/\partial I = n_2-2 n_4 I.
\label{n2_eff}
\end{equation}
Filamentation starts if self-focusing prevails over diffraction i.e if the beam power $P$ exceeds a critical power $P_{cr}$. For low intensities the self-focusing is defined by cubic Kerr effect only and $P_{cr} \propto 1/n_2$. However, the self-focusing leads to an increase of the intensity and the corresponding decrease of $ n_2^{(eff)}$. One can introduce the effective critical power  $P_{cr}^{(eff)}=P_{cr} n_2 / n_2^{(eff)}$. The self-focusing stops when $P_{cr}^{(eff)}=P$. Taking into account Eq.~(\ref{n2_eff}) this condition gives
{\color{purple}
\begin{equation}
  I= I_{sat}\left( 1-{P_{cr}}/{P}\right).
  \label{I_fil}
\end{equation}
}

 From the numerical polarizability for nitrogen shown in Fig.~\ref{al_vs_int} for the experimental pulse duration of 100 fs we find {\color{purple} the saturation intensity $I_{sat, N_2}^{(TDSE)} = 1.4$~TW/cm$^2$. From a similar calculation for oxygen we find $I_{sat, O_2}^{(TDSE)} = 1.1$~TW/cm$^2$. For air we use the average of these two values. The three-level model using the published properties of the nitrogen molecule (see Appendix E) gives $I_{sat, N_2}^{(analyt)} = 0.7$~TW/cm$^2$.}

 {\color{purple}
 The critical powers and the experimental beam  powers for different gases are calculated using the parameters presented in Table I. Using the values of saturation intensity described above, we calculate the intensities in filaments via Eq.~(\ref{I_fil}). The results are also presented in Table~I. Comparing them with the measured ones, we see that the TDSE-based filament intensities exceed the experimental ones by a factor of 1.4-2.1. However, our simulations reproduce the fact that in oxygen and air the filament intensities exceed that in nitrogen by a factor of approximately 3 and 1.5-2, respectively.   Moreover, the filament intensity found via the three-level model for nitrogen agrees with the experimental result (within the experimental error). }


Thus, in this paper, we study experimentally and numerically the filamentation of the 248 nm UV pulses with a duration of 100 fs in nitrogen, oxygen, and air at atmospheric pressure. We measure the intensity in the filaments and, in particular, find that, counterintuitively, the intensity in oxygen is approximately three times higher than in nitrogen. Moreover, our quantitative estimations of the free-electron densities in filaments show that these densities are orders of magnitude lower than those required to compensate for self-focusing. 
Via the numerical TDSE solution for model systems, we calculate the nonlinear refractive index in the intense UV field. The refractive index first grows linearly with the laser intensity due to the cubic Kerr-effect, then saturates, and finally decreases. The saturation intensity depends on the laser pulse duration (thus behaves non-perturbatively): for the longer pulses, saturation occurs at lower intensities. The experimentally observed threefold excess of the filament intensity in oxygen with respect to that in nitrogen is well reproduced in our calculations. 

Both experiments and simulations show that the contribution of photoelectrons to the saturation of the refractive index is negligible in the UV domain. The calculations allow us to attribute this saturation to the nonlinear transient excitation of the molecule. Such excitation is effective in the UV field because two or three photons are sufficient for resonant excitation of a molecule. This excitation reduces the refractive index due to the disappearance of the molecules in the ground state (having a positive linear refractive index), as well as due to the appearance of the excited molecules (having a negative linear refractive index in the UV domain). Clarifying the origin of the filamentation of the UV radiation in gases, our results allow not only the correct description of the filament properties but also open new perspectives to control these properties using resonant features of the UV interaction with different atmospheric gases.

\section*{Data availability statement}
Data supporting the findings of this article are openly available~\cite{data_UV_filamentation}.

\section*{Acknowledgment}
The authors acknowledge O. Kosareva for fruitful discussions and D. Mokrousova for technical assistance. 

\input{suppl.txt}

\bibliography{lit}

\end{document}

%% file: suppl.txt
\section*{Supplement} 
\subsection{Time-dependent Schr\"odinger equation numerical solution}

We solve numerically the 3D time-dependent Schr\"odinger equation (TDSE), atomic units are used throughout unless otherwise specified:
\begin{equation}
i\frac{\partial}{\partial t}\Psi(\textbf{r},t) = \left(-\frac{1}{2}\nabla^{2}+V(\textbf{r})+x \, E(t)\right)  \Psi(\textbf{r},t)
\label{TDSE}
\end{equation}
for a single-electron system with a binding potential $V(\textbf{r})$ in the laser field $E(t)$ linearly polarized in the x-direction. The method of the TDSE numerical solution is described in~\cite{Strelkov_2006}. The TDSE is solved on a spatial grid and the wave-function which reaches the boundary of the numerical box is absorbed by an additional imaginary component of the potential, see~\cite{Strelkov_dark_2023}. The numerical box is large enough to describe correctly the Rydberg states excited by the UV field, see below.

According to Ehrenfest theorem   
$$
 \ddot{x}(t) = F_{Coulomb}(t)-E(t)
$$
where $$x(t)=<\Psi(\textbf{r},t)|x|\Psi(\textbf{r},t)>$$ and $$F_{Coulomb}(t)=-<\Psi(\textbf{r},t)|\partial V/ \partial x|\Psi(\textbf{r},t)>.$$ For the Fourier components we have $x(\omega)=-F_{Coulomb}(\omega)/\omega^2+E(\omega)/\omega^2$. 

Polarizability is defined as $\alpha=d(\omega_0)/E(\omega_0)$, where $d(\omega_0)$ is the atomic dipole moment at the fundamental frequency $\omega_0$. Taking into account that $d =-x$ we have:
\begin{equation}
\alpha=\frac{F_{Coulomb}(\omega_0)}{E(\omega_0) \omega_0^2} - \frac{1}{\omega_0^2}.
\label{polarization}
\end{equation}

The latter equation is used for the numerical calculation of the polarizability. Note that when the electron finds itself in the continuum due to the photoionization it is absorbed when it reaches the boundary of the numerical box. However, the polarization calculated via Eq.~(\ref{polarization}) is still correct because the Coulomb force is zero for the electron that is far from the origin, so the absorption of this electron does not affect the Coulomb force. The second term on the right side of Eq.~(\ref{polarization}) (i.e. the free electron polarizability $\alpha_{el}=-1/\omega_0^2$) is constant and thus it is not affected by the absorption either.

In Fig.~2 for polarizability in the main text we also present the sum of the cubic Kerr and the contribution of free electrons. The latter is calculated as $(1-\rho)\alpha_{el}$, where $\rho=<\Psi(\textbf{r}, t=\tau/2)|\Psi(\textbf{r},t=\tau/2)>$ is the norm of the wave function at the center of the laser pulse (at time $t=\tau/2$). The norm decreases in time because when the electron finds itself in the continuum due to ionization it reaches the boundary of the numerical box and is absorbed. Note that under the conditions we used the free electron is absorbed rapidly, because its minimal kinetic energy (four laser photons minus the ionization energy) corresponds to a high velocity. Thus, $1-\rho$ is a very reasonable approximation of the ionization probability. 

The model potential is
\begin{equation}
\begin{array}{lcl}
  V(r)= \\ \\
  -V_0+\frac{2}{a(r)+b(r)+\sqrt{\left[  a(r)-b(r)\right]^2+\delta^2 \exp\left[ -\left(\frac{r}{R_0}\right)^2 \right]}},\\
\end{array}
\label{V_r}
\end{equation}

where $$a(r)=\frac{1}{V_{h}(r)},$$ $$b(r)=\frac{1}{V_{C}(r)},$$ $V_{h}$ is a harmonic potential: $$V_h(r)=\Omega_0^2r^2/2,$$ $V_C(r)$ is a (shifted) Coulomb potential: $$V_C(r)=V_0-1/r,$$ where
\begin{equation}
  V_0=I_p+3 \Omega_0/2,  
  \label{V_0}
\end{equation} 
$I_p$, $\Omega_0$, $\delta$, $R_0$ are constants. 

The potential $V(r)$ given by Eq.~(\ref{V_r}) has the following properties: for all $r$ (except for a narrow range, see below) the first term in the square root far exceeds the second, so the denominator is approximately  written as $a(r)+b(r)+|a(r)-b(r)|$. For small $r$ we have $a>b$, so the denominator equals $2a$ and therefore $V(r) \approx V_h -V_0$. Similarly, for large $r$ we find $V(r) \approx -1/r$. For a narrow range of $r$ we have $a\approx b$, so the second term under the square root is not negligible. This term provides a smooth switch from the harmonic to the Coulomb potential. The smoothness is controlled by the parameter $d$ and the distance for which the potential is almost Coulomb is defined by the parameter $R_0$. 

\begin{figure}
\includegraphics[width=0.35\textwidth]{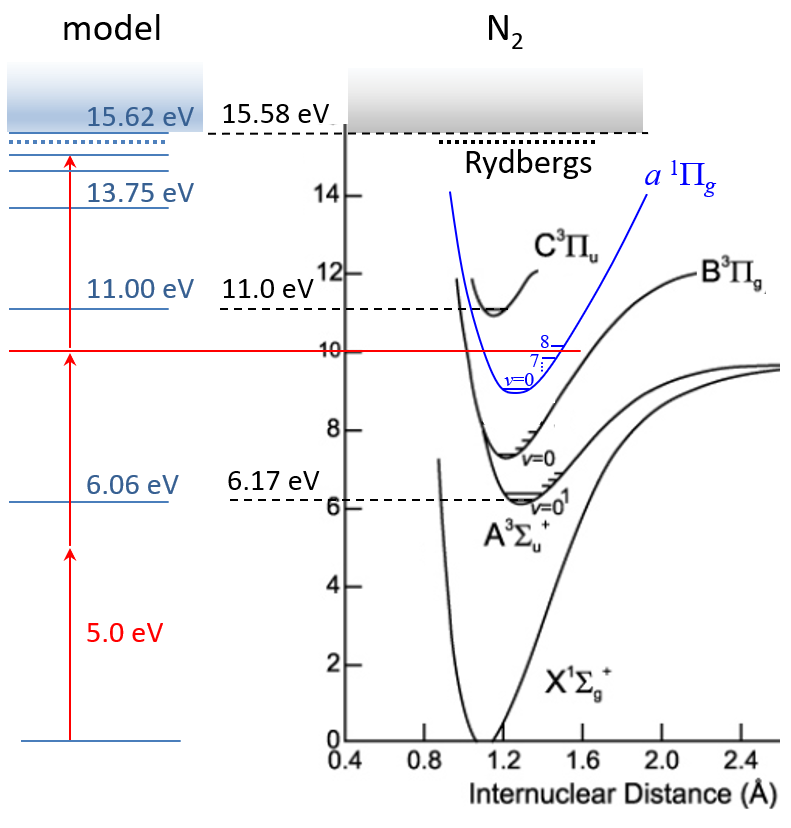}
\caption{The simplified level structure of the $N_2$ molecule (right) and the level structure of the model potential (left). The red arrows show the UV photons.}
\label{model_illustration}
\end{figure}

For the harmonic potential $V_h(r)-V_0$ the energy of the ground state is $-I_p$ (see Eq.~(\ref{V_0})) and the excitation energy of the first excited state is $\Omega_0$. This gives a very good analytical approximation for the energies of the two lowest states in the potential~(\ref{V_r}).  The energy of the second excited state is somewhat lower than $-I_p+2 \Omega_0$ due to the deviation of $V(r)$ from the harmonic law. Parameters $d$ and $R_0$ allow us to control its energy. Finally, finding numerically the actual energies of the ground states, we tune the parameters to construct the desired (see below) level structure for the three lowest levels.

\subsubsection*{Nitrogen molecule}
\textcolor{purple}{The values found to reproduce the relevant parameters of the nitrogen molecule} are  $I_p=0.581$, $\Omega_0=0.228$, $d=0.2$, $R_0=10$. Note that for large $r$ the potential $V(r)$ is close to the Coulomb one, so the Rydberg levels are also reproduced in our calculations.

Thus, choosing the parameters of the potential~(\ref{V_r}), we construct the level structure of the model potential that allows to mimic to some extent the interaction of the nitrogen molecule with the UV field, see~Fig.~\ref{model_illustration}. 

The excitation of the molecule with the UV photons can occur via several paths. This can be \textcolor{purple}{ the dipole-allowed two-photon resonant excitation from the ground state (the lowest state of the $X^1\Sigma_g^{+}$ term) to the $a \ ^1\Pi_g$ term, more details are presented in Appendix E. Other options can be the following sequential excitations: }

- the first photon excites from $X^1\Sigma_g^{+}$ to $A ^3\Sigma_u^{+}$ states (Vegard-Kaplan System~\cite{spectum_of_nitrogen}); the second photon excites from $A ^3\Sigma_u^{+}$ to  $B^3\Pi_g$ (First Positive System); 

-  the first photon excites from $X^1\Sigma_g^{+}$ to  $B^3\Pi_g$ (Wilkinson System); the second photon excites from $B^3\Pi_g$ to $C^3\Pi_u$ (Second positive System); the third photon excites from  $C^3\Pi_u$ to Rydbergs;

-  the first photon excites from $X^1\Sigma_g^{+}$ to $A \ ^3\Sigma_u^{+}$ (Vegard-Kaplan System); the second photon excites from $A \ ^3\Sigma_u^{+}$ to $E \ ^3\Sigma_g^{+}$ (Herman-Kaplan System); the third photon excites from $E \ ^3\Sigma_g^{+}$ to Rydbergs.

The energies of the first and second excited states in our model are close to the energies of the lowest vibration states of the $A^3\Sigma_u^{+}$ and $C^3 \Pi_u$ terms, respectively, see Fig.~\ref{model_illustration}. The properties of the Rydberg states are similar in the molecule and the atom. Finally, we reproduce the ionization potential which is 15.58~eV for the actual molecule and 15.62~eV for our model.

{
\color{purple}
\subsubsection*{Oxygen molecule}
The values found to reproduce the relevant parameters of the oxygen molecule are  $I_p=0.459$, $\Omega_0=0.27$, $d=0.2$, $R_0=10$. For these values of the parameters the model potential reproduces the ionization energy of the molecule and the frequency of the lowest dipole allowed transition (the lowest frequency of the Schumann-Runge system). 

}


\subsection{Time-dependent polarizability}
The time-dependent polarizability is calculated using a wavelet transform (namely, the Gabor transform) of the atomic response and the field. Namely, we calculate the masked response as:
\begin{equation}
\tilde d (t,t') = d(t') \exp \left[ -\left( \frac{t-t'}{\delta t}\right)^2\right],
\end{equation}
where we use $\delta t = 0.15 \tau$ and its spectrum as $\tilde d (t, \omega') = \int \tilde d (t,t') \exp(i \omega' t') dt ' $. Then we calculate the Gabor transform of the field $\tilde E (t, \omega')$ and the time-dependent polarizability as 
\begin{equation}
\tilde \alpha (t) = \tilde d (t, \omega_0)/ \tilde E (t, \omega_0).
\label{eq:Gabor}    
\end{equation}

\begin{figure}
\includegraphics[width=0.45\textwidth]{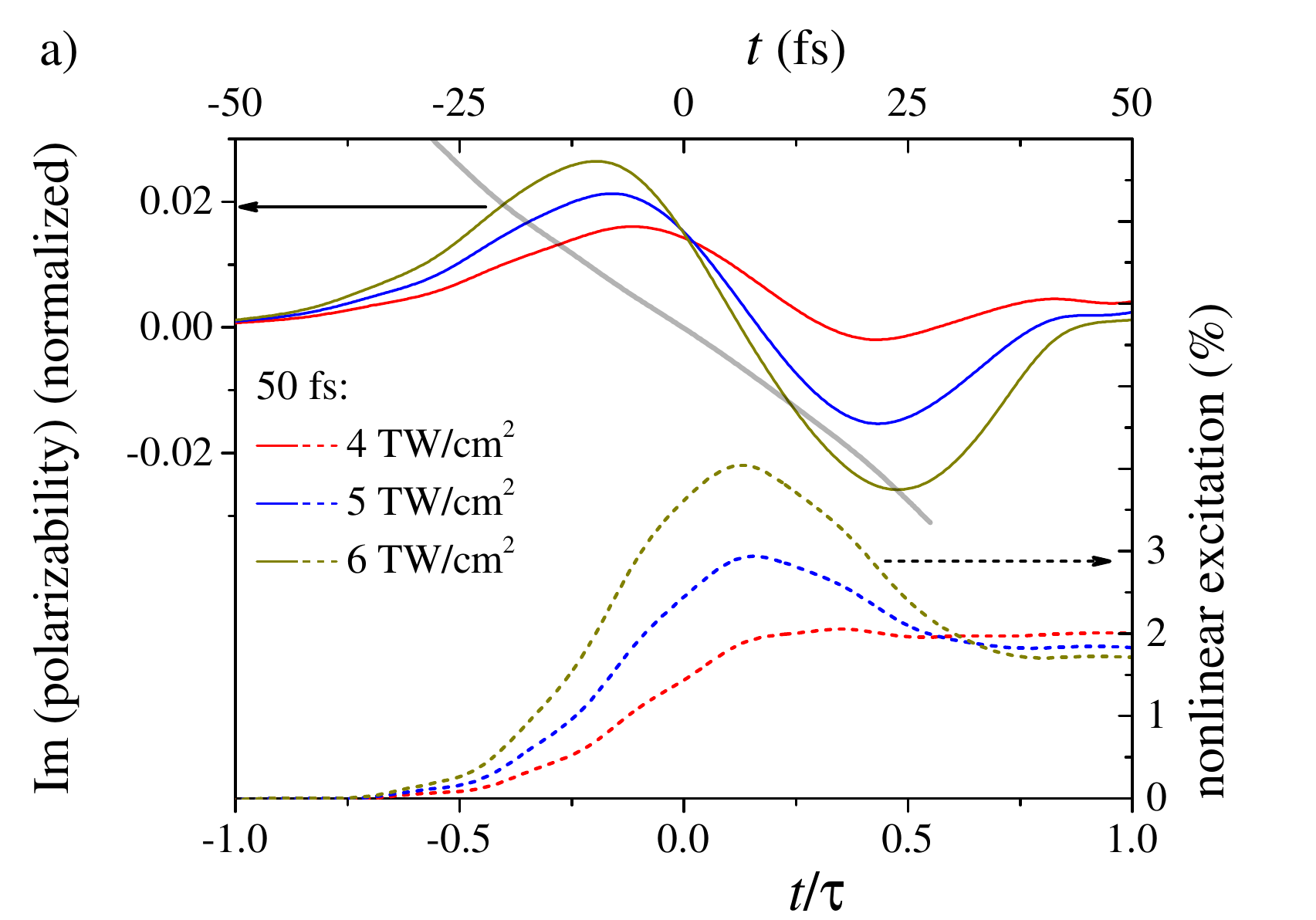}
\includegraphics[width=0.45\textwidth]{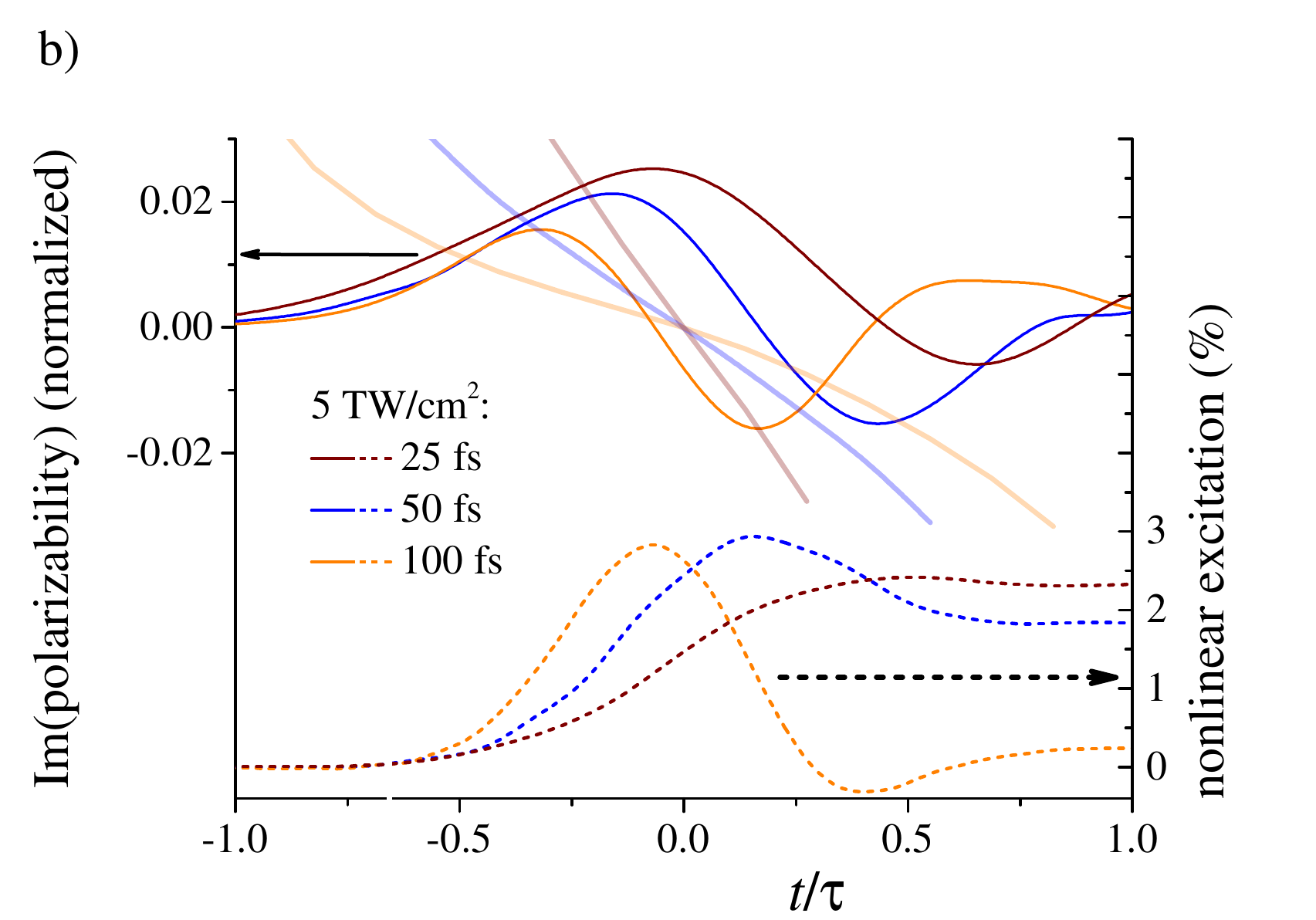}
\caption{The imaginary part of the polarizability (solid curves) and nonlinear excitation (dashed curves) as functions of time for different peak pulse intensities (a) and different pulse durations (b). The light thick curves present the polarizability in the weak field ${\rm Im} \left[\tilde \alpha^{(weak)} (t)\right ]/\alpha_0$, and the thin curves present nonlinear part of the polarizability ${\rm Im} \left[   \tilde  \alpha^{(nl)} (t) \right]/\alpha_0$ calculated via Eq.~(\ref{al_nl}).}
\label{Im_Gabor}
\end{figure}

\subsection{The nonlinear excitation and the imaginary part of the polarizability}
In the above subsections we discuss the properties of the real part of the polarizability. In this one we focus on its imaginary part. Fig.~\ref{Im_Gabor} presents its time evolution, namely, the value ${\rm Im} \left[ \tilde \alpha(t)\right]/\alpha_0 $. In the following we denote the polarizability calculated for low intensity $I^{(weak)}= 0.1$~TW/cm$^2$ as the linear polarizability $\tilde \alpha^{(weak)} (t)$. Unlike its real part, the imaginary part is not constant in time, see light thick curves in the figure. The linear phase variation of the complex polarizability in the weak field corresponds to the dispersion of the response. The non-linear polarizability is calculated as:
\begin{equation}
 \tilde  \alpha^{(nl)} (t)= \tilde \alpha (t)-\tilde \alpha^{(weak)} (t).
    \label{al_nl}
\end{equation}

We see that the imaginary part of the polarizability is much smaller than its real part, compare Fig.~2 in the main text and Fig.~\ref{Im_Gabor}. The linear and the non-linear contributions to the imaginary part of the polarizability are comparable, unlike the similar contributions to the real part. 

In Fig.~\ref{Im_Gabor} one can see that the imaginary part of the nonlinear polarization is proportional to the time derivative of the nonlinear excitation (this can be attributed to the fact that the field is absorbed to excite the system, so the imaginary part of the polarization is positive when the excitation increases). Therefore, the total absorption $ \int I(t) \ {\rm Im} [\tilde \alpha (t)] dt $ is proportional to the final excitation of the system.

{\color{purple}

The decrease of the nonlinear excitation with the intensity decrease after the peak of the laser pulse can be attributed to the properties of the Rabi oscillations. The Rabi oscillations in the populations of different excited states are to a large extent averaged in Fig. 3 and in Fig.~\ref{Im_Gabor}, because the figures present the total nonlinear excitation of many states. For every transition, the exited state population is proportional to the Rabi frequency squared, so it is proportional to the intensity (for a one-photon transition) or intensity squared (for a two-photon one), see Eq.~(\ref{W_t}) in Appendix E.  Thus, the populations of the majority of the excited states transiently grow with the laser intensity at the front of the pulse and then decrease with the pulse intensity decrease. Thus, the total contribution of these transitions to the nonlinear excitation adiabatically follows the intensity variation. However, for some transitions the generalized Rabi frequencies are low, so the periods of the Rabi oscillations are comparable to the pulse duration and the contributions of these transitions do not adiabatically follow the intensity variation; if also a detuning from the resonance for such transition is low, the transition contributes much to the total excitation. It makes the behavior of the total excitation more complex, in particular, can lead to some residual excitation remaining after the pulse passes.
}

\begin{figure}
\includegraphics[width=0.45\textwidth]{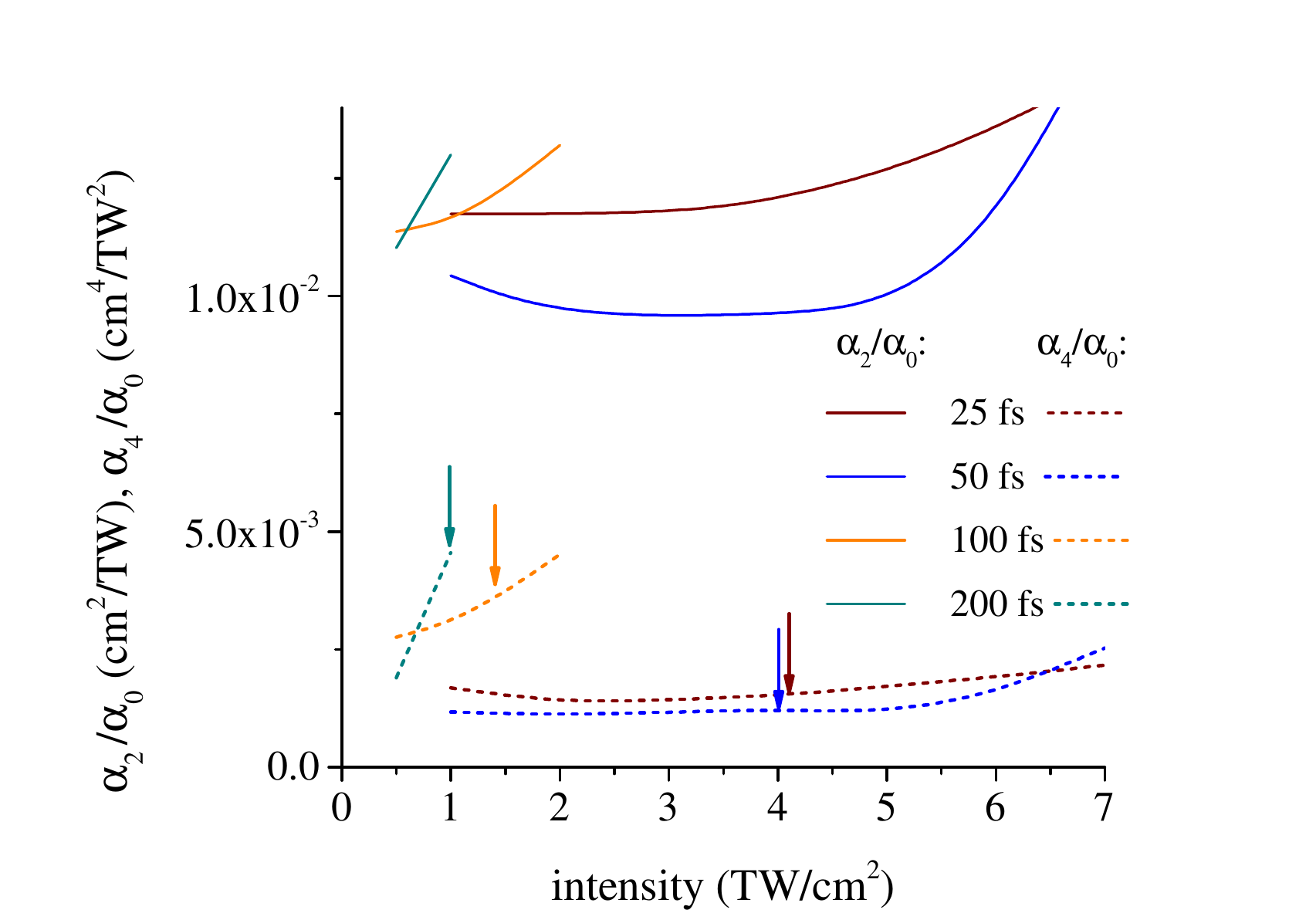}
\caption{The coefficients obtained approximating numerical polarizability shown in Fig.~2 in the main text with expansion~(\ref{alpha}), for different pulse durations. The arrows show the intensities where the dependence $\alpha(I)$ (for a given pulse duration) saturates. }
\label{al_series}
\end{figure}

\subsection{Polarizability expansion in a power series of intensity}

For a given pulse duration, one can expand the real part of the numerical polarizability as a power series of the intensity as
\begin{equation}
 \alpha(I)=\alpha_0+\alpha_2 I-\alpha_4I^2.
 \label{alpha}
\end{equation}
Certainly, such expansion assumes that coefficients $\alpha_{2,4}$ do not depend on $I$. However, numerically these coefficients are found to provide a good approximation of 
$\alpha(I)$ within a certain intensity interval from $0$ to $I_{max}$. Fig.~\ref{al_series} presents $\alpha_{2,4}$ (found using the least squares method) for the several pulse durations as functions of $I_{max}$. One can see that for the pulse durations 100~fs and shorter, the coefficients weakly depend on $I_{max}$ up to the intensities where the dependence $\alpha(I)$ saturates. \textcolor{purple}{So using the expansion (\ref{alpha}) and thus the expansion (4) in the main text is reasonable.} For the pulse duration of 200~fs $\alpha_4$ rapidly varies even for an intensity below saturation. This means that for this pulse duration one has to consider higher terms in Eq.~(\ref{alpha}) for an accurate description of the $\alpha(I)$ dependence.


{
\color{purple}

\subsection{Estimation of the molecular excitation probability and the refractive index saturation intensity}

\textcolor{purple}{In this subsection we estimate the probability of the dipole-allowed two-photon excitation of the nitrogen molecule from the ground state (the lowest state of the $X^1\Sigma_g^{+}$ term) to the $a \ ^1\Pi_g$ term, see Fig.~\ref{model_illustration}.  Similar two-photon transitions were investigated in the studies of the (2+2) REMPI of nitrogen~\cite{Trickl_1993, Mori_1999, McGuire_2014, Gardner_2019}.} 

\textcolor{purple}{We study this excitation using the three-level model. We choose the $b \ ^1\Pi_u$ term as an intermediate state because:}

\textcolor{purple}{- both the $X^1\Sigma_g^{+} - b \ ^1\Pi_u$ transitions (Birge-Hopfield I system) and the  $b \ ^1\Pi_u - a \ ^1\Pi_g$ transitions (Gaydon-Herman system)  are  dipole-allowed;}

\textcolor{purple}{- although the excitation energy for the $b \ ^1\Pi_u$ states is essentially higher than $\omega_0$ (see below), it is smaller than the excitation energies for the other terms which can be excited via the allowed transitions from the ground state, because the $b \ ^1\Pi_u$ term is the {\it lowest} term for which the transitions from the ground state are allowed.} 

\textcolor{purple}{Below we denote the parameters of the transition from the ground to the intermediate state by the index "1i" and those of the transition from the intermediate state to the resonantly-excited (i. e. $a \ ^1\Pi_g$) state by the index "i2". The population of the latter state as a function of time is:}
\begin{equation}
\textcolor{purple}{p_{ex}=\frac{\Omega^2}{\Omega_{gen}^2} \sin^2 \frac{\Omega_{gen} t}{2},}
\label{W_t}
\end{equation}
\textcolor{purple}{where $\Omega$ is the two-photon Rabi frequency,
\begin{equation}
\textcolor{purple}
{
\Omega_{gen}=\sqrt{\Omega^2+\delta^2}
}
\label{Rabi_gen}
\end{equation}
is the generalized two-photon Rabi frequency, $\delta$ is the detuning of the resonantly excited state from the two-photon resonance. Several vibrational states of the $a \ ^1\Pi_g$ term are close to the resonance. In particular, for $v=7$ the energy is $E_7=9.92$~eV and for $v=8$ the energy is $E_8=10.11$~eV~\cite{Trickl_1993}, which is close to the total energy of the two photons, namely  $2 \omega_0=10.00$~eV. We use the energy of the closest state to calculate the detuning: $\delta=|E_7-2\omega_0|$}.    

\textcolor{purple}{The two-photon Rabi frequency is:}
\begin{equation}
\textcolor{purple}{\Omega=\frac{\Omega_{1i} \Omega_{i2}}{2 \Delta},}
\label{Rabi2}
\end{equation}
\textcolor{purple}{where $\Delta$ is the detuning of the intermediate state. We use the energy of the lowest $b \ ^1\Pi_u$ state $E_{1i}=12.5$ eV to calculate this detuning $\Delta=|E_{1i}-\omega_0|$. $\Omega_{1i, i2}$ are the Rabi frequencies of the transitions in the three level-system, $\Omega_{1i, i2}= r_{1i, i2} E_0$, $r_{1i, i2}$ are the dipole moment matrix elements, $E_0$ is the field amplitude.}

\textcolor{purple}{From Eq.~(\ref{Rabi_gen}) we find that the period of the Rabi oscillation is about 50~fs. Taking into account that the pulse duration is longer, below we deal with the time averaged population $\bar p_{ex}$, substituting $<\sin^2 (\Omega_{gen} t/2)>$ in Eq.~(\ref{W_t}) by $1/2$. Moreover, to take into account that not only the vibrational state with $v=7$, but the neighboring vibrational states of the $a \ ^1\Pi_g$ term are populated as well, we multiply the population by a factor of 2. So below we use $\bar p_{ex} = {\Omega^2}/{\Omega_{gen}^2} $. From Eqs.~(\ref{Rabi_gen}),~(\ref{Rabi2}) we have:}

\begin{equation}
    \bar p_{ex} = \frac{r^2_{1i}r^2_{i2} E_0^4}{4 \delta^2 \Delta^2}
    \label{W_av}
\end{equation}

As we mentioned in the main text, the change in the refractive index due to excitation can be described as the disappearance of a particle with polarizability of the molecule in the ground state $\alpha_0$ and the simultaneous appearance of a particle with polarizability of the molecule in the excited state $\alpha_{ex}$. In the main text we assumed that the latter equals the polarizability of a free electron $\alpha_{el}=-1/\omega_0^2$, because an electron in the excited molecule is not bound very tightly (to be more precise, the frequencies of the transitions describing further excitation are smaller than $\omega_0$), so its behavior in the high-frequency field is similar to that of the free electron. To check the accuracy of this assumption, we calculate numerically the polarizability of the molecule in the $a \ ^1\Pi_g$ state. One-photon transitions to the lower states are forbidden, the closest upper state to which the transition is allowed is the lowest $b \ ^1\Pi_u$ state. So we use the potential~(\ref{V_r}) with $I_p=0.206$, $\Omega_0=0.11$, $d=0.2$, $R=10.0$ providing the ionization energy of $15.6-10=5.6$~eV and the first excited state energy $12.6-10=2.6$~eV. We solve the TDSE for such a system in a weak UV field and find that its polarizability is close to that of the free electron, namely 
\begin{equation}
 \alpha_{ex}=-a/\omega_0^2,
 \label{al_ex}
\end{equation}
where $a=1.4$ for $\omega_0=5$~eV.

Thus, the change in the polarizability due to the excitation is $\bar p_{ex} (-\alpha_0+\alpha_{ex})$. Using the Lorentz-Lorenz equation, which connects the refractive index and the polarizability for gases as: 
\begin{equation}
n=1+\alpha 2\pi N    
\end{equation}
(where $N$ is the gas density), we find for the change in the refractive index: 
\begin{equation}
    \Delta n_{ex}=\bar p_{ex} (-\alpha_0+\alpha_{ex}) 2\pi N,
\end{equation}
 The intensity $I_{sat}$ for which the refractive index saturates is found from the condition:$$ \left. \frac{\partial \Delta n_{ex} }{\partial I} \right |_{I=I_{sat}}=-n_2.$$ From the latter Eq. and Eqs.~(\ref{W_av}),~(\ref{al_ex}) we find:
\begin{equation}
  I_{sat}=\frac{ n_2 }{\left [ \left (n_0-1 \right )+2\pi N a/\omega_0^2 \right ]}  \frac{2 \delta^2 \Delta^2}{r_{1i}^2 r_{i2}^2}.
  \label{I_sat}
\end{equation}

The oscillator strength for the $X^1\Sigma_g^{+} - b \ ^1\Pi_u$ transition found in~\cite{Rose_1973, Glushkov_1992} is $f_{1i} = 0.3$. This gives $r_{1i} = 1.0$. Note that one can roughly estimate the matrix element $r_{eff}$ of a single transition with the energy $E_{int}$ providing the actual refractive index $n_0$ of the nitrogen for the used UV frequency. This gives $r_{eff}=2.2$. As expected,  $r_{1i}<r_{eff}$, but closeness of the two values shows that the Birge-Hopfield I transitions considerably define the non-resonant one-photon interaction of the used UV with the molecule in the ground state and thus justifies choosing $b \ ^1\Pi_u$ as an intermediate state in our model. We assume that the dipole matrix element characterizing $b \ ^1\Pi_u - a \ ^1\Pi_g$ transitions should be close to that characterizing $X^1\Sigma_g^{+} - b \ ^1\Pi_u$ ones, so we use $r_{i2}=1.0$.

Substituting these values in Eq.~(\ref{W_av}) we find $\bar p_{ex}=2.8 \, \, 10^{-4}$ for the UV intensity of 1~TW/cm$^2$ and $\bar p_{ex}=7 \, \, 10^{-3}$ for 5~TW/cm$^2$. This  value is compatible with the results of our TDSE simulations, see Fig. 3 in the main text.

Using $n_2$ value for nitrogen (see the main text) from Eq.~(\ref{I_sat}) we find  $I_{sat, N_2}^{(analyt)}=0.7$~TW/cm$^2$. This is twice lower than the saturation intensity found in the TDSE simulations $I_{sat, N_2}^{(TDSE)} =1.4$~TW/cm$^2$, see Fig 2. So, the filament intensity estimated within the three-level model is $0.17$~TW/cm$^2$, thus close to the experimental result, see table I. 

\subsection{Saturation of the refractive index for different UV frequencies}

We made TDSE calculations of the saturation intensity (the intensity for which the refractive index stops growing with the laser intensity) for 100 fs pulses with different frequencies. The results are presented in Fig.~\ref{Fig_I_sat}. For the laser photon energy in the range from 4.5 eV to 5.5 eV the saturation intensity changes in a complex way, but remains comparable to the value found for 5 eV. For higher photon energies strong UV absorption makes the saturation intensity calculation senseless. For lower photon energies the saturation intensity suddenly grows up to values leading to high ionization probability, and for these conditions plasma provides the dominant contribution to the Kerr effect compensation. Thus, the TDSE calculations show that in the range from 4.5 eV to 5.5 eV the predominant compensation mechanism is molecular excitation.

Our estimation based on the three-level model (Appendix E) also does not predict a dramatic change of the saturation intensity within a certain range of the laser frequencies. This is because several  vibrational states provide comparable contributions to the molecular excitation (for pulse durations of about 100 fs and shorter, considered here), so the variation of the frequency detuning changes the contributions of the different vibrational states but does not significantly change the total excitation probability. The two-photon resonant excitation via the nitrogen molecule terms considered in our model is impossible for the photon frequencies below 4.3 eV (a half of the lowest state of the  $a ^1 \Pi_g$ term), so this is the low-frequency limit of the excitation-dominated regime in our three-level model. 

Note, that using more complicated models (for instance, considering more molecular states) could lead to further decrease of this limit frequency. Moreover, the frequency range of the excitation-dominated regime depends on the pulse duration, and the comprehensive investigation of this issue is an outlook for our future studies.

\begin{figure}
\includegraphics[width=0.45\textwidth]{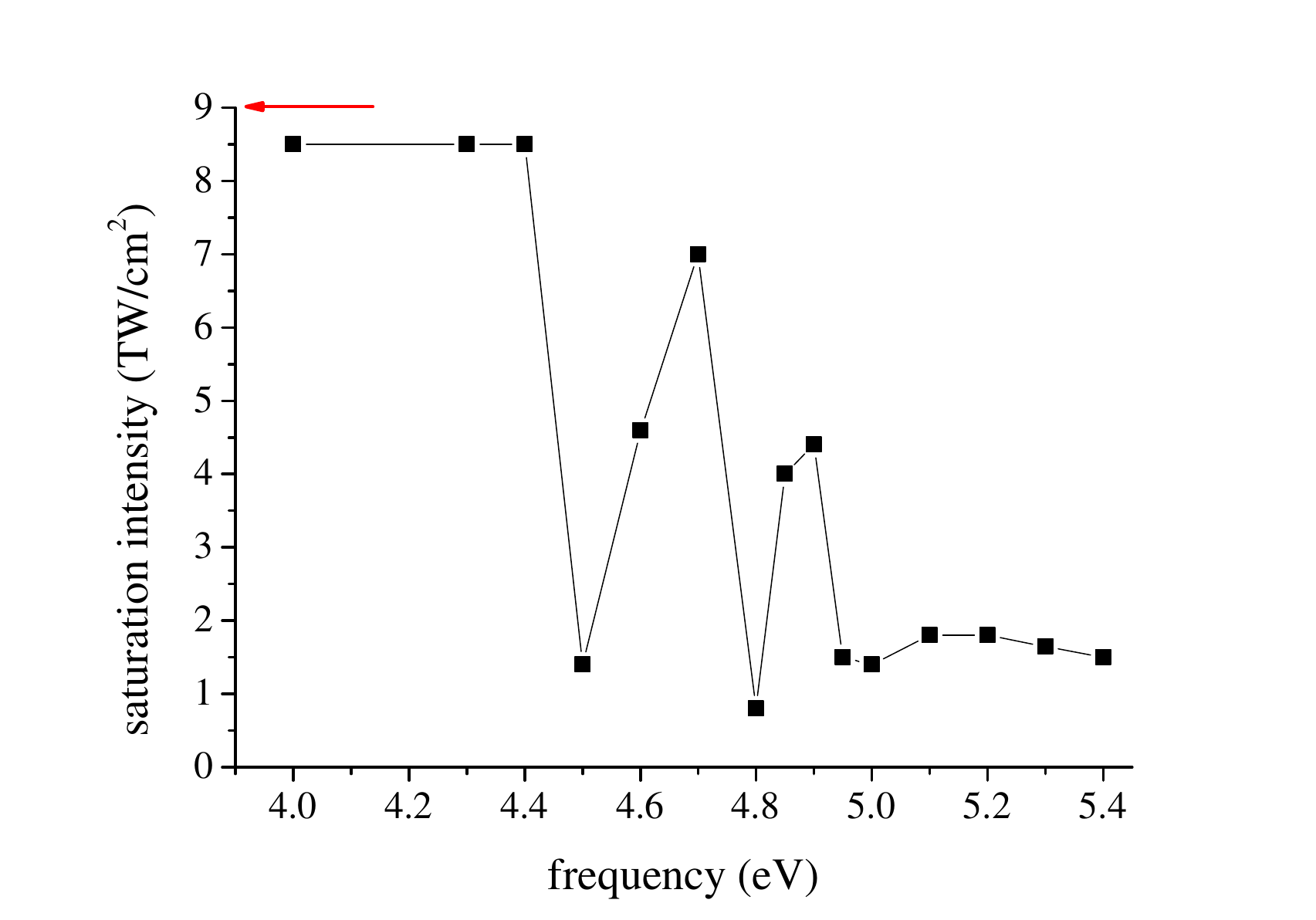}
\caption{\color{purple}{The saturation intensity calculated using numerical TDSE solution for different UV field frequencies. The pulse duration is 100 fs. The red arrow shows the intensity providing photoionization rate which is sufficient for the Kerr-effect compensation.}  }
\label{Fig_I_sat}
\end{figure}

}